\documentclass[twocolumn,showpacs,amsmath,amssymb,10pt,aps]{revtex4}

\usepackage{graphicx,amsmath,bbm,mathrsfs,amssymb,times,color,tikz-cd}
\usetikzlibrary{arrows,decorations.pathmorphing,backgrounds,positioning,fit,petri,calc,shapes.misc,decorations.markings}
\tikzset{degil/.style={
                decoration={markings,
                mark= at position 0.5 with {
                \node[transform shape] (tempnode) {$\backslash$};
                }
                },
                postaction={decorate}
} }

\def\beq{\begin{equation}}
\def\eeq{\end{equation}}
\def\beqq{\begin{eqnarray}}
\def\eeqq{\end{eqnarray}}

\newtheorem{theorem}{Theorem}
\newtheorem{corollary}{Corollary}

\newtheorem{remark}{Remark}
\newtheorem{proposition}{Proposition}
\newcommand{\qed}{\hfill $\Box$\medskip}
\newcommand{\proof}{\noindent{\bf Proof: }}

\begin{document}

\title{Tensor power of dynamical maps and P- vs. CP-divisibility}

\author{Fabio Benatti}
\affiliation{Dipartimento di Fisica, Universit\`a degli Studi di
Trieste, I-34151 Trieste, Italy}
\affiliation{Istituto Nazionale di Fisica Nucleare, Sezione di
Trieste, I-34151 Trieste, Italy}
\author{Dariusz Chru\'sci\'nski}
\affiliation{Institute of Physics,
Faculty of Physics, Astronomy and Informatics \\  Nicolaus Copernicus University,
Grudzi{a}dzka 5/7, 87--100 Torun, Poland}
\author{Sergey Filippov}
\affiliation{Moscow Institute of Physics and Technology,
Institutskii Per. 9, Dolgoprudny, Moscow Region 141700, Russia}


\begin{abstract}
The are several non-equivalent notions of Markovian quantum
evolution. In this paper we show that the one based on the
so-called CP-divisibility of the corresponding dynamical map
enjoys the following stability property: the dynamical map
$\Lambda_t$ is CP-divisible iff the second tensor power
$\Lambda_t\otimes\Lambda_t$ is CP-divisible as well. Moreover, the
P-divisibility of the map $\Lambda_t\otimes\Lambda_t$ is
equivalent to the CP-divisibility of the map $\Lambda_t$.
Interestingly, the latter property is no longer true if we replace
the P-divisibility of $\Lambda_t\otimes\Lambda_t$ by simple
positivity and the CP-divisibility of $\Lambda_t$ by complete
positivity. That is, unlike when $\Lambda_t$ has a
time-independent generator, positivity of
$\Lambda_t\otimes\Lambda_t$ does not imply complete positivity of
$\Lambda_t$.
\end{abstract}


\maketitle

\section{Introduction}

Usually, the non-unitary, dissipative time-evolution of an open
quantum system $S$, that we take as a finite-level system, for
sake of simplicity, is approximated by dynamical maps $\Lambda_t$
that are constrained to be completely positive \cite{AL,Book}.
Namely, if the system $S$ is initially statistically coupled to an
inert, non-evolving copy of it, the dynamics $\Lambda_t\otimes{\rm
id}$ of $S+S$ must be positive, that is it must map all possible
initial states of $S+S$ into density matrices, thus guaranteeing
the positivity of their spectrum at all times. Otherwise, there
surely exist entangled states of $S+S$ whose spectrum acquires
negative eigenvalues that cannot then be interpreted as
probabilities~\cite{AL}. However, in view of the generic and
uncontrollable character of the ancilla, such a motivation for the
necessity  of complete positivity is scarcely physically
palatable, above all because of the ensuing constraints which, in
the case of a Markovian dynamical semigroup,
$\Lambda_t=\exp(t\,L)$, are embodied by the celebrated
Gorini-Kossakowski-Sudarshan-Lindblad form of the generator
$L$~\cite{GKS,L}
\begin{equation}\label{GKSL}
L[\rho]  =  -i [H,\rho] + \frac 12 \sum_\alpha \gamma_\alpha
\left( [V_\alpha, \rho
 V_\alpha^\dagger] + [V_\alpha \rho,  V_\alpha^\dagger]  \right)
\end{equation}

\noindent with positive decoherence/dissipation rates
$\gamma_\alpha > 0$.

In~\cite{BFR} a more physical point of view was presented, whereby
both systems are embedded into a same environment thus undergoing
the same dissipative Markovian time-evolution, $\Lambda_t$. In
this case, the ancilla is not out of practical control and not
inert, the compound system $S+S$ dynamics being described by
$\Lambda_t\otimes\Lambda_t$. Physical consistency then demands
that the latter map be positive so to exclude the appearance of
negative probabilities in the spectrum of time-evolving states of
$S+S$. It turns out that the complete positivity of $\Lambda_t =
\exp(tL)$ is equivalent to positivity of the tensor product
$\Lambda_t\otimes\Lambda_t$. Hence positivity of $e^{tL}\otimes
e^{tL}$ implies the generator $L$ to be of the Lindblad form
(\ref{GKSL}).

Consider now the time-local master equation
\begin{equation}
\label{D1} \frac{{\rm d}}{{\rm d}t}\Lambda_t = L_t\,\Lambda_t\
,\qquad \Lambda_{t=0} = \mathbb{I}\ ,
\end{equation}

\noindent governed by time-local generator $L_t$. One may wonder
wether a similar result holds for the solution $\Lambda_t$; that
is, is it true that positivity of $\Lambda_t \otimes \Lambda_t$
implies complete positivity of $\Lambda_t$? In this paper we show
that this result is no longer true for general $L_t$ (in the next
Section  we provide a concrete counterexample of random unitary
qubit evolution). However, it holds for Markovian dynamical maps.
Hence, violation of the above implication may be considered as
another witness of non-Markovianity of $\Lambda_t$. Non-Markovian
quantum evolutions have recently been extensively analyzed (see
\cite{rev1,rev2,rev3} for recent reviews and the collection of
papers in \cite{rev4}). There are several non-equivalent
definitions of Markovian evolution \cite{rev1,rev2,rev3}. In this
paper we adopt the one based on the concept of divisibility.
Recall that $\Lambda_t$ is divisible if $\Lambda_t = \Lambda_{t,s}
\Lambda_s$ for all $t\geq s\geq 0$. Moreover, a divisible map
$\Lambda_t$ is

\begin{itemize}
\item CP-divisible if $\Lambda_{t,s}$ is completely positive,

\item P-divisible if $\Lambda_{t,s}$ is positive.
\end{itemize}

Notice that, if $\Lambda_t$ is CP-divisible,  then
$\Lambda_t=\Lambda_{t,s=0}$ is completely positive for all
$t\geq0$. Analogously, if $\Lambda_t$ is P-divisible then
$\Lambda_t$ is at least positive for all $t\geq 0$. We call the
quantum evolution Markovian iff the corresponding dynamical map
$\Lambda_t$ is CP-divisible \cite{RHP,Wolf-Isert,JPB,PRL-Sabrina}.
In the following, we show that the P-divisibility of $\Lambda_t
\otimes \Lambda_t$ implies that $\Lambda_t$ is CP-divisible, hence
corresponding to a Markovian evolution. Clearly, the
CP-divisibility of $\Lambda_t$ implies the CP-divisibility of
$\Lambda_t \otimes \Lambda_t$; then, on the level of the tensor
product $\Lambda_t \otimes \Lambda_t$ P- and CP-divisibility are
equivalent.  This proves that the notion of Markovianity based on
the concept of CP-divisibility is stable with respect to replacing
$\Lambda_t$ with the tensor product $\Lambda_t \otimes \Lambda_t$.


\section{Positive, not Completely Positive Qubit Dynamics}

In this Section we construct a positive (but not completely
positive) map $\Lambda_t$ such that $\Lambda_t\otimes \Lambda_t$
is positive. Consider the following qubit time-local generator
\begin{equation}\label{Lt}
L_t[\rho] = \frac{\alpha}{2} \sum_{k=1}^3\gamma_k(t) ( \sigma_k
\rho \sigma_k - \rho) \ ,
\end{equation}

\noindent where $\sigma_j$, $j=1,2,3$ are the Pauli matrices,
$\gamma_1(t)=\gamma_2(t)=1$, and $\gamma_3(t) = -\tanh(t)$. The
parameter $\alpha > 0$ controls the property of the corresponding
map $\Lambda_t$. For $\alpha=1$ this generator was already
considered in \cite{Erika} as an example of so-called {\em
eternal} non-Markovian evolution (see also \cite{Nina}).

\begin{proposition} The corresponding map $\Lambda_t$ is
\label{prop1}
\begin{itemize}

\item positive  for all $\alpha > 0$,

\item completely positive iff $\alpha \geq 1$.

\end{itemize}

\end{proposition}

\proof Let us represent a qubit density matrix by a Bloch vector $\mathbf{r}=(r_1,r_2,r_3)$ such that
\begin{equation}
\label{DM}
\varrho=\frac{1}{2}\Big(\mathbb{I}+\sum_{j=1}^3r_j\sigma_j\Big)\
,\qquad r_j\in\mathbb{R}\ ,\quad \sum_{j=1}^3r_j^2\leq 1\ .
\end{equation}

\noindent When complemented with the identity matrix $\sigma_0=
\mathbb{I}$, the matrices $\sigma_\mu$, $\mu=0,1,2,3$,  are
eigenvectors of $L_t$,
\begin{equation}
\label{mateig}
  L_t[\sigma_\mu] = \lambda_\mu(t) \sigma_\mu ,
\end{equation}
with eigenvalues
$$
\lambda_0(t)=0\ , \ \lambda_1(t)=\lambda_2(t) = \alpha [\tanh(t) -1]\ , \ \lambda_3(t)=-2\alpha\ .
$$

\noindent One then readily gets the following time-evolution
equations for the Bloch vector components of $r_j(t)$ of
$\varrho_t=\Lambda_t[\varrho]$,
$$
\frac{{\rm d}r_{1,2}(t)}{{\rm d}t}=\alpha \frac{\tanh(t)-1}{2}\,r_{1,2}(t)
\ ,\quad
\frac{{\rm d}r_{3}(t)}{{\rm d}t}=-2\alpha r_3(t)\ ,
$$

\noindent so that a straightforward integration yields
\begin{equation}
\label{DM1} \Lambda_t [\varrho] = \frac{1}{2}\Big(\mathbb{I} +
{e}^{-\alpha t}\cosh^\alpha(t)
(r_1\sigma_1+r_2\sigma_3)+{e}^{-2\alpha t}r_3 \sigma_3\Big)\ .
\end{equation}

The map $\varrho \rightarrow \Lambda_t[\varrho]$ is positive since
${e}^{-\alpha t}[\cosh(t)]^\alpha \leq 1$ for $t \geq 0$. In order
to analyze the complete positivity of $\Lambda_t$, let us observe
that its action can be recast in the form
\begin{equation}
\label{KSF}
\Lambda_t=\sum_{\mu=0}^3 p_\mu(t)S_\mu\ ,
\end{equation}

\noindent where $S_\mu[\varrho] = \sigma_\mu \varrho \sigma_\mu$
and the parameters $p_\mu(t)$ are~\cite{CW}:
\begin{eqnarray}\label{ppp}
  p_0(t) &=& \frac 14 \left( 1 + 2e^{-\alpha t} \cosh^\alpha(t) + e^{-2\alpha t} \right) , \nonumber \\
  p_1(t) &=& p_2(t) =  \frac 14 \left( 1  { -} e^{-2\alpha t} \right), \\
  p_3(t) &=& \frac 14 \left( 1 {-} 2e^{-\alpha t} \cosh^\alpha(t) {+} e^{-2\alpha t} \right) \nonumber .
\end{eqnarray}

\noindent Clearly, $\Lambda_t$ is CP iff (\ref{KSF}) corresponds
to a Kraus representation, that is, iff $p_\mu(t) \geq 0$ for all
$t\geq 0$. The only nontrivial condition $p_3(t) \geq0$ is
equivalent to
\begin{equation}\label{COSH}
\cosh(\alpha t)\geq \cosh^\alpha(t)\ ,
\end{equation}

\noindent which is satisfied iff $\alpha \geq 1$. Indeed, $f(t) =
\ln \cosh t$ has a positive second derivative and is thus convex.
Hence, since $f(0)=0$, for any $0\leq\alpha \leq 1$ one has
$$
f\left(\alpha t + (1- \alpha) \cdot 0\right) \leq \alpha f(t) + (1-\alpha)\,f(0)\ ,
$$
so that~\eqref{COSH} is violated. On the other hand, if $\alpha\geq 1$,
$$
f(t)=f\left(\frac{1}{\alpha} (\alpha t) + \left(1- \frac{1}{\alpha}\right) \cdot 0\right) \leq \frac{1}{\alpha} f(\alpha t)\ ,
$$
whence~\eqref{COSH} follows. \qed

As briefly outlined in the Introduction, when $\Lambda_t$ has a
time independent generator, the lack of complete positivity of
$\Lambda_t$ and thus of positivity of $\Lambda_t\otimes{\rm
id}_2$, is often not regarded as a compelling argument in favour
of complete positivity. This is so because envisioning possible
initial quantum correlations of the system of interest with an
ancilla, another generic qubit in the present case, otherwise
completely independent and inert,  looks more as a mathematical
request than a necessary physical constraint. Moreover, the
consequences of such an abstract motivation are none the less
physically quite relevant. Indeed, despite the fact that it is
perfectly well behaved on single qubit states, a time-evolution as
$\Lambda_t$ in~\eqref{DM1} is ruled out as physically inconsistent
because it is $\Lambda_t\otimes{\rm id}_2$ which is physically
ill-defined: indeed, it cannot keep positive all possible
initially entangled states.

However, if instead of a generic, uncontrollable ancilla, one
considers another system under the same physical conditions of an
open system as the previous one and non-interacting with the
former, then the dynamics of the compound system becomes
$\Lambda_t\otimes\Lambda_t$. Unlike $\Lambda_t\otimes{\rm id}_2$,
$\Lambda_t\otimes\Lambda_t$  is physically more tenable and
physical consistency demands it to be positive. In~\cite{BFR} it
was proved that, in the Markovian case when the time-local
generator $L_t$ is in fact time-independent, $L_t=L$,
$\Lambda_t\otimes\Lambda_t$ is positive if and only if $\Lambda_t$
is completely positive.

We now show, by means of a counterexample, that this conclusion
does not hold in the more general setting represented by the
master equation~\eqref{D1}. The main technical tool is the
following result proved in Proposition 4 of~\cite{Filippov}.

\begin{proposition}
\label{prop1}
If $\Lambda_t$ is a linear map on the algebra
$M_2(\mathbb{C})$ of $2\times 2$ matrices and $\Lambda^2_t$ is
completely positive, then $\Lambda_t\otimes\Lambda_t$ is positive.
\end{proposition}

\begin{proposition}
\label{lem4} The maps $\Lambda_t$ in \eqref{DM1} satisfy the
following property:  $\Lambda_t\otimes\Lambda_t$ is positive for
all $\alpha \geq \frac 12$.
\end{proposition}

\proof We show that for $\alpha \geq\frac 12$ the map
$\Lambda_t^2$ is completely positive and hence, due to Proposition
\ref{prop1}, the tensor product $\Lambda_t\otimes\Lambda_t$ is
positive. Using the Pauli matrix algebra, one reduces the product
$S_{\mu}S_{\nu}$ to the action of single $S_\lambda$ and finds
\begin{equation}
\Lambda^2_t=\sum_{\mu=0}^3 q_\mu(t)S_\mu\ ,
\end{equation}

\noindent with parameters
\begin{eqnarray*}
  q_0(t) &=& \frac 14 \left( 1 + 2e^{-2\alpha t} \cosh^{2\alpha}(t) + e^{-4\alpha t} \right) ,  \\
  q_1(t) &=& q_2(t) =  \frac 14 \left( 1  {-} e^{-4\alpha t} \right) , \\
  q_3(t) &=& \frac 14 \left( 1 {-} 2e^{-2\alpha t} \cosh^{2\alpha}(t) { +} e^{-4\alpha t} \right) ,
\end{eqnarray*}

\noindent which differ from (\ref{ppp}) by an obvious replacement
$\alpha \rightarrow 2\alpha$. Then, if $\alpha \geq\frac 12$ one
has $q_\mu(t)\geq0$. \qed

\begin{remark}
Putting together Proposition~\ref{prop1} and
Proposition~\ref{lem4}, it follows that for $\alpha \in [\frac
12,1)$ the map $\Lambda_t$ is positive but not completely positive
whereas the tensor product  $\Lambda_t\otimes\Lambda_t$ is
positive. This way we provided a counterexample to the naive
expectation that the property -- $\Lambda_t$ is completely
positive iff $\Lambda_t\otimes\Lambda_t$ is positive-- that holds
for time-independent generators~\cite{BFR}, might also hold for
general master equations of the form~\eqref{D1}. Thus, in general,
the relations between the (complete) positivity of the maps
$\Lambda_t$ and the (complete) positivity of the maps
$\Lambda_t\otimes\Lambda_t$ can be summarized by the following
diagram:
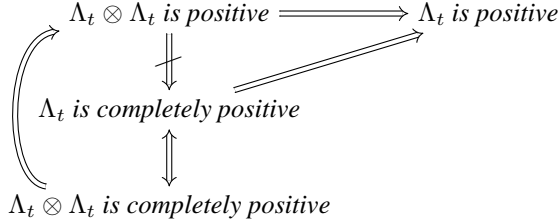
\[
\begin{tikzcd}
\Lambda_t \otimes \Lambda_t \text{ is positive} \arrow[degil,Rightarrow]{d}{}\arrow[Rightarrow]{r}{} & \Lambda_t \text{ is positive}  \\
\Lambda_t \text{ is completely positive} \arrow[Leftrightarrow]{d}{}\arrow[Rightarrow]{ur}{} \\
\Lambda_t \otimes \Lambda_t \text{ is completely positive}
\arrow[Rightarrow,to path={..controls +(-2.2,0.35) and
+(-2.2,-0.35).. (\tikztotarget)}]{uu}{}
\end{tikzcd}
\]
\end{remark}

\begin{remark}
Interestingly, Proposition~\ref{lem4} also provides a
counterexample to another naive expectation that if $L_t$
generates completely positive dynamical maps $\Lambda_t$, then the
rescaled $cL_t$ with $c >0$  does the same. Noticeably, the model
of random unitary evolution (\ref{KSF}) was recently used for
describing the effective dynamics of disordered quantum systems
\cite{Chahan}.
\end{remark}


\section{P- and CP-divisibility}

While the positivity of $\Lambda_t\otimes\Lambda_t$ does not in
general require $\Lambda_t$ to be completely positive when the
generator of $\Lambda_t$ is time-dependent, in this section, we
shall instead show that the P-divisibility of
$\Lambda_t\otimes\Lambda_t$ implies the CP-divisibility of
$\Lambda_t$. Indeed,  the following result holds whose proof is an
adaptation from~\cite{BFR}.

\begin{theorem}
\label{thD1}
The one-parameter family $\{\Lambda_t\}_{t\geq 0}$ on the state space of a $d$-level system is CP-divisible if and only if $\{\Lambda_t\otimes\Lambda_t\}_{t\geq 0}$ is P-divisible.
\end{theorem}

\proof

The scheme of the proof is as follows:
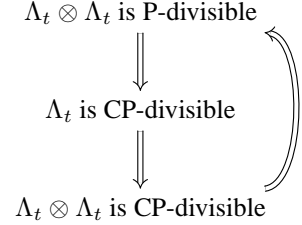
\[
\begin{tikzcd}[arrows=Rightarrow]
\Lambda_t \otimes \Lambda_t \text{ is P-divisible} \arrow{d}{} \\
\Lambda_t \text{ is CP-divisible} \arrow{d}{} \\
\Lambda_t \otimes \Lambda_t \text{ is CP-divisible} \arrow[to
path={..controls +(2.2,0.35) and +(2.2,-0.35)..
(\tikztotarget)}]{uu}{}
\end{tikzcd}
\]

Because of linearity and trace-preservation, the action of the
local time-dependent generator $L_t$ on a state $\varrho$ can
always be written in the form
$$
L_t[\varrho]=-i[H_t,\varrho]+\sum_{i,j=1}^{d^2-1}C_{ij}(t)\,\left(
F_i\varrho F_j^\dag-\frac{1}{2}\left\{F_j^\dag F_i,\varrho\right\}\right)\ ,
$$

\noindent with respect to an orthonormal Hilbert-Schmidt basis of
$d^2-1$ $d\times d$ traceless, matrices such that ${\rm
Tr}(F_j^\dag F_i)=\delta_{ij}$ complemented with
$F_{d^2}=1/\sqrt{d}$. The only consistency request on the
$(d^2-1)\times(d^2-1)$ matrix $C(t) =[C_{ij}(t)]$ is that it be
Hermitian.

The P-divisibility of $\Lambda_t\otimes\Lambda_t$ implies that the
maps
$$
\Lambda_{t,s}\otimes\Lambda_{t,s}=\mathcal{T}\exp\left(\int_s^t{\rm d}u\,\Big(L_u\otimes {\rm id}+{\rm id}\otimes L_u\Big)\right)\ ,
$$

\noindent with ${\rm id}$ the identity operation, are positive for
all $t\geq s\geq0$; namely that
$$
\Lambda_{t,s}\otimes\Lambda_{t,s}
[\vert\psi\rangle\langle\psi\vert] \geq 0\ ,\quad \forall t\geq
s\geq0\ ,\forall
\vert\psi\rangle\in\mathbb{C}^d\otimes\mathbb{C}^d\ .
$$

\noindent Choosing $\vert\phi\rangle\perp\vert\psi\rangle$ and
expanding $\Lambda_{t,s}\otimes\Lambda_{t,s}$, one obtains, up to
first order in $t-s\geq 0$ with $s\geq0$ fixed,
\begin{widetext}
\begin{eqnarray*}
0\leq\langle\phi\vert\Lambda_{t,s}\otimes\Lambda_{t,s}[\vert\psi\rangle\langle\psi]\vert\phi\rangle\simeq\Delta_{t-s} &:=&(t-s)\,\Big(\langle\phi\vert L_s\otimes{\rm id}[\vert\psi\rangle\langle\psi\vert]\vert\phi\rangle+
\langle\phi\vert{\rm id}\otimes L_s[\vert\psi\rangle\langle\psi\vert]\vert\phi\rangle\Big)
\\
&=&(t-s)\sum_{i,j=1}^{d^2-1}C_{ij}(s)\Big(
\langle\phi\vert F_i\otimes 1\,\vert\psi\rangle\langle\psi\vert F_j^\dag\otimes 1\vert\phi\rangle+
\langle\phi\vert 1\otimes F_i\vert\psi\rangle\langle\psi\vert 1\otimes F_j^\dag\vert\phi\rangle
\Big)\ .
\end{eqnarray*}
\end{widetext}

\noindent Fixing an orthonormal basis in $\mathbb{C}^d$ and
regrouping the $d^2$ components $\psi_{ab}$ of $\vert\psi\rangle$
and $\phi_{ab}$ of $\vert\phi\rangle$ into $d\times d$ matrices
$\Psi=[\psi_{ab}]$ and $\Phi=[\phi_{ab}]$, one can set $u^*_i :=
\langle\phi\vert F_i\otimes 1\vert\psi\rangle$, $v^*_i :=
\langle\phi\vert 1\otimes F_i\vert\psi\rangle$ and write
\[
u^*_i =\sum_{a,b,c=1}^d\phi_{ab}^*F_i^{ac}\psi_{cb} =
{\rm Tr}\Big(\Psi\,\Phi^\dag\,F_i\Big)\ ,
\]
\[
v^*_i =\sum_{a,b,d=1}^d\phi_{ab}^*F_i^{bd}\psi_{ad} =
{\rm Tr}\Big(\Big(\Phi^\dag\,\Psi\Big)^{tr}\,F_i\Big)\ ,
\]

\noindent so that
$$
\Delta_{t-s}=(t-s)\Big(\langle u\vert C(s)\vert u\rangle+\langle v\vert C(s)\vert v\rangle
\Big)\ , 
$$
where $\vert u\rangle$ and $\vert v\rangle$ are
$(d^2-1)$-dimensional vectors with components $u_i$ and $v_i$,
respectively. Choosing $\Phi^\dag=U$, $\Psi=M\,U^{-1}$, $M,U\in
M_{d^2-1}(\mathbb{C})$, $U$ being the similarity matrix such that
$M^{tr}=U\,M\,U^{-1}$ (such a matrix $U$ always exsists), one
finds $\Psi\Phi^\dag=M$, $\Phi^\dag\Psi=M^{tr}$, whence $\vert
u\rangle=\vert v\rangle$.

The orthogonality of $\vert\psi\rangle$ and $\vert\phi\rangle$
amounts to asking that ${\rm Tr}(\Phi^\dag\Psi)={\rm Tr}(M)=0$;
then, tracelessness is the only constraint $M$ must fulfil.
Therefore, varying $M$ one can achieve any $\vert u\rangle$ in
$\mathbb{C}^{d^2-1}$. The positivity of
$\Lambda_{t,s}\otimes\Lambda_{t,s}$ asks for
$$
\Delta_{t-s}=2(t-s)\,\langle u\vert C(s)\,\vert u\rangle\geq 0\qquad \forall\,
\vert u\rangle\in\mathbb{C}^{d^2-1}\ ,
$$

\noindent which  in turn yields the positive semi-definiteness of
the coefficient matrix $C(s)\in M_{d^2-1}(\mathbb{C})$. Such a
condition is sufficient for the complete positivity of the maps
$\Lambda_{t,s}$. Then, the one-parameter family
$\{\Lambda_t\}_{t\geq0}$ is CP-divisible.

Vice-versa, if $\{\Lambda_t\}_{t\geq0}$ is CP-divisible, then the
maps $\Lambda_{t,s}$, $t\geq s\geq 0$, are completely positive, as
well as  the tensor products $\Lambda_{t,s}\otimes\Lambda_{t,s}$
so that the one-parameter family
$\{\Lambda_t\otimes\Lambda_t\}_{t\geq 0}$ is CP- and thus
P-divisible. \qed

One has the following straightforward implications
\begin{corollary}
\label{corD1}
$\Lambda_t\otimes\Lambda_t$ is P-divisible if and
only if $\Lambda_t\otimes\Lambda_t$ is CP-divisible.
\end{corollary}

\begin{corollary}
\label{corD2}
$\Lambda_t$ is Markovian if and only if
$\Lambda_t\otimes\Lambda_t$ is Markovian.
\end{corollary}

The model studied in the previous Section provides the following
intriguing observation.

\begin{corollary}
\label{corD3} For $\alpha\in [\frac 12,1)$ the maps $\Lambda_t$ in~\eqref{DM1} are such that

\begin{itemize}

\item $\Lambda_t$ is positive but not completely positive

\item $\Lambda_{t,s}$ is positive for $t > s\geq 0$ which means that $\Lambda_t$ is P-divisible

\item $\Lambda_t\otimes\Lambda_t$ is positive,

\item $\Lambda_{t,s}\otimes\Lambda_{t,s}$ cannot be positive for all $t> s\geq 0$.

\end{itemize}

\end{corollary}

\proof The one-parameter family of the maps $\Lambda_t$
in~\eqref{DM1} is P-divisible: this follows from a result
in~\cite{CW} together with the fact that $\gamma_i(t) +
\gamma_j(t) \geq 0$ for $i \neq j$. Then, the maps $\Lambda_{t,s}$
are positive for all $t\geq s\geq 0$. If the tensor product maps
$\Lambda_{t,s}\otimes\Lambda_{t,s}$ were also positive for all
$t\geq s\geq 0$, then the one-parameter family
$\Lambda_t\otimes\Lambda_t$ would be P-divisible and hence
CP-divisible, according to Corollary~\ref{corD2}. Then, the maps
$\Lambda_{t,s}$ would be completely positive for all $t\geq s\geq
0$ contradicting the fact that $\Lambda_t:=\Lambda_{t,s=0}$ are
positive, but not completely positive. \qed

\begin{remark}
The previous corollary shows that, unlike the notion of
Markovianity based on CP-divisibility, the one based on the
vanishing back-flow of information~\cite{BLP} is not stable with
respect to the tensor product. Let us recall that following
\cite{BLP}  one can define the information flow by means of
\begin{equation}\label{}
  \sigma(\varrho_1,\varrho_2;t) = \frac{d}{dt} || \Lambda_t[\varrho_1-\varrho_2]||_1 ,
\end{equation}

\noindent where $\varrho_1$ and $\varrho_2$ are arbitrary density
operators of the system. According to~\cite{BLP}, the evolution is
defined Markovian if $ \sigma(\varrho_1,\varrho_2;t) \leq 0$ for
any $\varrho_1$, $\varrho_2$ and $t \geq 0$. Whenever  $
\sigma(\varrho_1,\varrho_2;t) > 0$, the two density matrices
become more distinguishable and this fact is identified as
information flowing from the environment into the system which
provides a clear sign of memory effects. Now, for time-evolutions
generated by~\eqref{Lt}, the definition of Markovianity as absence
of back-flow of information coincides with P-divisibility
\cite{CW,PRL-Sabrina}. Hence, Proposition~\ref{prop1} provides an
example of dynamical maps $\Lambda_t$ with vanishing back-flow of
information, such that their tensor product $\Lambda_t \otimes
\Lambda_t$ none the less gives rise to nontrivial back-flow of
information.
\end{remark}


\section{Conclusions}

In this paper we have discussed the consequences of asking that a
one-parameter family of dynamical maps $\Lambda_t$ consist of
$\Lambda_t$ such that $\Lambda_t\otimes\Lambda_t$ be positive.
Unlike when $\Lambda_t=\exp(t\,L)$, in the case $\Lambda_t$ is
generated by a time-local master equation, the positivity of
$\Lambda_t\otimes\Lambda_t$ does not enforce the complete
positivity of $\Lambda_t$. It is however the P-divisibility of
$\Lambda_t\otimes\Lambda_t$ that implies the CP-divisibility of
$\Lambda_t$. There follow interesting connections between the
P-divisibility, which defines classical Markovian evolutions, and
the CP-divisibility, which defines Markovianity in the quantum
case. We have also revealed an interesting phenomenon of
superactivation of the back-flow of information, namely, there
exist dynamical maps $\Lambda_t$ with vanishing flow of
information from the environment into the system such that the
second tensor power $\Lambda_t \otimes \Lambda_t$ nevertheless
induces non-zero back-flow of information.


\section*{Acknowledgements}
D.C. was partially supported by the National Science Centre project
2015/17/B/ST2/02026. S.F. thanks the Russian Science Foundation
for support under project No. 16-11-00084.



\begin{thebibliography}{999}
\bibitem{AL}
R. Alicki, K. Lendi, Quantum Dynamical Semigroups and Applications
(Springer, Berlin, 1987).

\bibitem{Book}
H.-P. Breuer and F. Petruccione, The Theory of Open Quantum
Systems (Oxford University Press, Oxford, 2007).

\bibitem{GKS}
A. Gorini, A. Kossakowski, and E.C.G. Sudarshan, J. Math. Phys.
{\bf 17}, 821 (1976).

\bibitem{L}
G. Lindblad, Comm. Math. Phys. {\bf 48}, 119 (1976).

\bibitem{BFR}
F. Benatti, R. Floreanini, and  R. Romano, J. Phys. A: Math. Gen.
\textbf{35}, L351 (2002).

\bibitem{rev1}
\'A. Rivas, S.F. Huelga, and M.B. Plenio, Rep. Prog. Phys.
{\bf77}, 094001 (2014).

\bibitem{rev2}
H.-P. Breuer, E.-M. Laine, J. Piilo, B. Vacchini, Rev. Mod.
Phys. {\bf 88}, 021002 (2016).

\bibitem{rev3}
I. de Vega and D. Alonso, {\em Dynamics of non-Markovian open
quantum systems}, arXiv:1511.06994 (to appear in Reviews of Modern
Physics).

\bibitem{rev4}
F. Benatti, R. Floreanini, and G. Scholes, J. Phys. B: At. Mol.
Opt. Phys. {\bf 45}, 150201 (2012).

\bibitem{RHP}
\'A. Rivas, S.F. Huelga, and M.B. Plenio, Phys. Rev. Lett. {\bf
105}, 050403 (2010).

\bibitem{Wolf-Isert}
M. M. Wolf, J. Eisert, T. S. Cubitt and J. I. Cirac, Phys. Rev.
Lett. \textbf{101}, 150402 (2008).

\bibitem{JPB}
D. Chru\'sci\'nski and A. Kossakowski, J. Phys. B: At. Mol. Opt.
Phys. {\bf 45}  154002 (2012).

\bibitem{PRL-Sabrina}
D. Chru\'sci\'nski and S. Maniscalco, Phys. Rev. Lett. {\bf 112},
120404 (2014).

\bibitem{Filippov}
S. N. Filippov and K. Yu. Magadov, {\em Positive tensor products
of qubit maps and 2-tensor-stable positive qubit maps},
arXiv:1604.01716v2.

\bibitem{CW}
D. Chru\'sci\'nski and F. A. Wudarski, Phys. Lett. A \textbf{377},
1425 (2013).

\bibitem{Filip-PRA}
D. Chru\'sci\'nski and F. A. Wudarski,  Phys. Rev. A {\bf 91},
012104 (2015).

\bibitem{Chahan}
C. M. Kropf, C. Gneiting, and A. Buchleitner, Phys. Rev. X {\bf
6}, 031023 (2016).

\bibitem{Nina}
N. Megier, D. Chru\'sci\'nski, J. Piilo, and W. T. Strunz, {\em
Non-Markovianity and the physics of memory: Be careful with master
equations}, arXiv:1608.07125.

\bibitem{BLP}
H.-P. Breuer, E.-M. Laine, J. Piilo, Phys. Rev. Lett. {\bf 103},
210401 (2009).

\bibitem{Erika}
E. Andersson, J. D. Cresser, and M. J. W. Hall,  Phys. Rev. A {\bf
89}, 042120 (2014).

\end{thebibliography}
\end{document}